\documentstyle[preprint,aps]{revtex}
\title{Exact, finite, and Hermitian fermion-pair-to-boson mapping}
\author{Calvin W.~Johnson and Joseph N.~Ginocchio}
\address{Theoretical Division, Los Alamos National Laboratory, Los Alamos, NM
87545}

\begin{document}
\draft

\maketitle
\begin{abstract}
\noindent
We discuss a general, exact (in that matrix elements are preserved) mapping
of fermion pairs to bosons, and find a simple factorization of the boson
representation of fermion operators.  This leads to boson Hamiltonians that
are Hermitian and finite, with no more than two-body operators if the fermion
Hamiltonian has at most two-body operators, and one-body boson transition
operators if the fermion transition operator is one-body.
\end{abstract}


Pairwise correlations are often important in describing the physics of
many-fermion systems.
The classic paradigm is the BCS theory of superconductivity\cite{BCS},
where the wavefunction is dominated
by Cooper pairs which have electrons coupled up to zero linear momentum and
spin; these
boson-like pairs condense into a coherent wavefunction.
Another example is the phenomenological Interacting Boson
Model (IBM) for nuclei\cite{IBM},
where many states and transition amplitudes are
successfully described using only $s$- and $d$- (angular momentum
$J=0,2$) bosons,
which represent coherent nucleon pairs.  In both cases
the large number of fermion degrees of freedom are well modelled by only a
few boson degrees of freedom.  On the other hand, however, despite some
forays by Otsuka et al.\ \cite{OAI}, a rigorous microscopic basis for such
phenomenological models is lacking.

The basic problem is to represent the underlying fermion dynamics and
statistics with a boson image amenable to approximation and numerical
calculation. Considerable effort has gone into mapping fermion pairs into
bosons \cite{Klein,Ring:Schuck}.  However
 these mappings typically suffer from a variety of defects.  Most, such as
the Belyaev-Zelevinskii \cite{BZ} and Marumori \cite{Maru} mappings give
rise to boson Hamiltonians with infinite expansions, that is, $N$-body
terms with $N \rightarrow \infty$.
Convergence is slow even
when ``collective'' fermion pairs are used. Finite  but non-Hermitian boson
Hamiltonians have also been derived \cite{Dyson,Rowe}.

In this Letter, using an alternate approach, we show that the infinite
expansion boson Hamiltonian
obtained from the exact mapping of fermion matrix elements to boson matrix
elements can be factorized into a finite, Hermitian boson Hamiltonian times a
norm operator, and it is the norm operator which has an
infinite boson expansion.

Consider a fermion space with $2\Omega$ single-particle states, and
a fermion Hamiltonian $\hat{H}$. The general problem is to solve the
fermion eigenvalue equation
\begin{equation}
\hat{H}\left | \Psi_p \right \rangle =
E_p \left | \Psi_p \right \rangle,
\label{eigeneqn}
\end{equation}
find transition amplitudes between eigenstates, and so on.  To do this we
require a many-body basis.
Often the basis set for many-fermion wavefunctions are Slater determinants,
antisymmeterized products of single-fermion wavefunctions which we can write as
 products of the Fock creation operators $a^{\dagger}_j, j = 1, \cdot \cdot
\cdot, 2 \Omega$ on the vacuum $a^{\dagger}_{i_{1}} \cdot \cdot \cdot
a^{\dagger}_{i_{n}} \left | 0 \right \rangle $ for $n$ fermions.  These states
span the antisymmetric
irreducible representation of the unitary group in
$2 \Omega$ dimensions, ${\rm SU}(2\Omega)$.
But for an even number of fermions one can just as well
construct states from $N = n/2$ fermion pair creation operators,
\begin{equation}
 \left | \psi_\beta \right \rangle
= \stackrel{N}{\prod_{m=1}} \hat{A}^\dagger_{\beta_m} \left | 0 \right \rangle,
\label{WfnDefn}
\end{equation}
with
\begin{equation}
\hat{A}^\dagger_\beta \equiv  { 1 \over \sqrt{2}} \sum_{ij}
\left ( {\bf A}^\dagger_\beta \right )_{ij}
{a}^\dagger_i {a}^\dagger_j.
\nonumber
\end{equation}
We always choose the  $\Omega(2\Omega-1)$ matrices ${\bf A}_\beta$ to be
antisymmetric to
preserve the underlying fermion statistics, and we choose for the
normalization the trace $ {\rm tr \,}  \left ( {\bf A}_\alpha {\bf
A}^\dagger_\beta \right )
= \delta_{\alpha \beta}$.
For this Letter we  represent generic one- and two-body operators by
\begin{equation}
\hat{T} \equiv
\sum_{ij} T_{ij} {a}^\dagger_i {a}_j, \, \, \, \,
\hat{V} \equiv \sum_{\delta \lambda} \left \langle \delta
\right | V \left |  \lambda \right \rangle
\hat{A}^\dagger_\delta \hat{A}_\lambda.
\end{equation}

We begin with the straightforward mapping to boson states
\begin{equation}
 \left | \psi_\beta \right \rangle
\rightarrow  \left | \phi_\beta \right )
= \stackrel{N}{\prod_{m=1}} {b}^\dagger_{\beta_m} \left | 0 \right ),
\label{BosonMap}
\end{equation}
where the ${b}^\dagger$ are boson creation operators.  In conjunction
with this simple mapping of states we construct boson operators that follow
the philosophy of Marumori \cite{Maru} and preserve matrix elements, for
example introducing boson operators $\hat{\cal T}_B$, $\hat{\cal V}_B$,
and most importantly $\hat{\cal H}_B$ such that
$ \left ( \phi_\alpha \right | \hat{\cal T}_B \left | \phi_\beta \right )
= \left \langle \psi_\alpha \right| \hat{T} \left | \psi_\beta \right \rangle$,
$
\left ( \phi_\alpha \right | \hat{\cal V}_B \left | \phi_\beta \right )
= \left \langle \psi_\alpha \right| \hat{V} \left | \psi_\beta \right \rangle,
$
and
$ \left ( \phi_\alpha \right | \hat{\cal H}_B \left | \phi_\beta \right )
= \left \langle \psi_\alpha \right| \hat{H} \left | \psi_\beta \right \rangle$.
In addition, because wavefunctions of the form (\ref{WfnDefn}) do not form
an orthonormal set,  we construct the norm operator $\hat{\cal N}_B$ with
the property
$ \left ( \phi_\alpha \right | \hat{\cal N}_B \left | \phi_\beta \right )
= \left \langle \psi_\alpha \right. \left | \psi_\beta \right \rangle$.
While the mapping (\ref{BosonMap}) is the usual Marumori mapping, this
boson representation of fermion operators is different, stemming from the
fact that  Marumori does not have an explicit norm operator.
With this mapping the fermion eigenvalue equation (\ref{eigeneqn}) becomes
a generalized (because of the norm) boson eigvenvalue equation
\begin{equation}
\hat{\cal H}_B \left | \Phi_p \right ) =
E_p   \hat{\cal N}_B \left | \Phi_p \right ).
\label{BosonEigen1}
\end{equation}
Because we have defined our boson operators so as to preserve matrix elements,
the original energy spectrum of (\ref{eigeneqn}) is found in
(\ref{BosonEigen1}).  However, because the boson space is much larger than
the original fermion space, (\ref{BosonEigen1}) also has additional
spurious states that do not correspond to physical fermion states.  These
by construction have zero eigenvalue.

When one constructs the norm operator \cite{JG} one finds it can be
conveniently expressed in terms of the  $k$th order Casimir operators of
the unitary group ${\rm SU}(2\Omega)$, $\hat{C}_k = 2^k \,{\rm tr \,} \left (
\sum_{\sigma \tau}
{b}^\dagger_\sigma {b}_\tau {\bf A}_\sigma {\bf A}^\dagger_\tau
\right)^k$,
\begin{equation}
\hat{\cal N}_B
= \, \colon \exp \left (
\sum_{k=2}^\infty
{ (-1)^{k-1} \over k}
\hat{C}_k \right ) \colon
\label{NormCasimirRep}
\end{equation}
where the colons `:' refer to normal-ordering of the boson operators.
Expanding (\ref{NormCasimirRep}) in normal order one obtains
the form \cite{JG}
\begin{equation}
\hat{\cal N}_B
= 1 + \sum_{\ell = 2}^\infty
\sum_{ \sigma \tau}
w_\ell^0 \left ( \left \{ \sigma \right \} ;
\left \{ \tau \right \} \right)
\prod_{m = 1}^\ell {b}^\dagger_{\sigma_m}
\prod_{n = 1}^\ell {b}_{\tau_n}.
\label{NormBosonRep}
\end{equation}
The $\ell$-body boson terms embody the fact that
fermion pair creation and annihilation operators do not have the
same commutation relations as do boson operators, and act to enforce the
Pauli principle.

The operators $\hat{\cal T}_B, \hat{\cal V}_B$ are also complicated many-body
operators similar in form to (\ref{NormBosonRep}), though upon examination
one can write them compactly using the norm operator.
For example,
\begin{equation}
\hat{\cal T}_B
 =  \colon \sum_{k = 1}^\infty (-1)^{k-1} 2^k \, {\rm tr \, }
\left [ \left ( {\bf T A}_\tau^\dagger {\bf A}_\sigma \right )
b^\dagger_\sigma b_\tau
\left ( \sum_{\alpha \beta} {\bf A}_\alpha {\bf A}_\beta^\dagger
b^\dagger_\alpha b_\beta \right )^{k-1} \right ]
\hat{ \cal N}_B \colon ,
\label{FullOneBody}
\end{equation}
where $({\bf T})_{ij}$ $= T_{ij}$, and
similarly for $\hat{\cal V}_B$ \cite{JG}.
In general these boson
operators do not have good convergence properties, so that truncation of
(\ref{BosonEigen1}), (\ref{FullOneBody}) as written  is problematic.

A key result of this Letter, as suggested by the explicit form of
(\ref{FullOneBody}), is that these operators factor in a
simple way:
$\hat{\cal T}_B = \hat{\cal N}_B \hat{T}_B =  \hat{T}_B \hat{\cal N}_B$ and
$\hat{\cal V}_B = \hat{\cal N}_B \hat{V}_B =  \hat{V}_B \hat{\cal N}_B$,
where the factored operators $\hat{T}_B, \, \hat{V}_B$, which we term  the
boson images of  $\hat{T}, \, \hat{V}$. have
simple forms.  For example,  a one-body fermion operator has a one-body boson
image
\begin{equation}
 \hat{T}_B =
2\sum_{\sigma \tau}
{\rm tr \,}
\left ( {\bf T A}_{\tau}^\dagger
{\bf A}_\sigma \right)
{b}^\dagger_{\sigma} {b}_\tau.
\end{equation}
To prove this factorization, one puts $\hat{T}_B \hat{\cal N}_B$ into
normal order:
\begin{equation}
2\sum_{\sigma \tau}
{\rm tr \,}
\left ( {\bf T A}_{\tau}^\dagger
{\bf A}_\sigma \right)
\left \{
{b}^\dagger_{\sigma} \hat{\cal N}_B
{b}_\tau
- {b}^\dagger_{\sigma} \left [ \hat{\cal N}_B,
{b}_\tau \right ] \right \} .
\end{equation}
The first term is the $k=1$ term from (\ref{FullOneBody}) and the
second term, the commutation of the norm operator with $b_\tau$,
gives rise to the $k > 1$ terms \cite{JG}.
Similarly,
\begin{equation}
\hat{V}_B  = \sum_{\delta \lambda} \left \langle \delta  \right |
V \left | \lambda \right \rangle \left[
{b}^\dagger_\delta {b}_\lambda
+ 2 \sum_{\sigma \sigma^\prime} \sum_{\tau \tau^\prime}
{\rm tr \,}
\left (
{\bf A}_\sigma {\bf A}^\dagger_\delta
{\bf A}_{\sigma^\prime} {\bf A}^\dagger_\tau
{\bf A}_\lambda {\bf A}^\dagger_{\tau^\prime}
\right )
{b}^\dagger_\sigma {b}^\dagger_{\sigma^\prime}
{b}_\tau {b}_{\tau^\prime}\right]
\end{equation}
and in general one can find a image Hamiltonian
\begin{equation}
\hat{H}_B = \hat{T}_B + \hat{V}_B.
\end{equation}
  Most of the complexity resides in the norm  operator and norm matrix and the
slow convergence of many-body terms that arise in other boson
mappings most likely do so because the norm is not completely factored out.

While we can demonstrate this factorization using the explicit forms for
$\hat{\cal T}_B$, $\hat{\cal N}_B$, etc., \cite{JG},
one can arrive  at this result directly. For example,  the action of a
one-body operator $\hat{T}$ on a fermion state $\left | \psi_\beta \right
\rangle$
is to linearly replace one fermion creation operator with another, or in
the fermion-pair
representation (\ref{WfnDefn}) , one fermion-pair creation operator with
another, and so schematically
$\hat{T}\left | \psi_\beta \right \rangle \rightarrow$ linear combination of
$\left | \psi_\gamma \right \rangle$.
This replacement is easily represented by bosons and it is straightforward
to show that in fact
\begin{equation}
\left \langle \psi_\alpha \right |
\hat{T}
\left | \psi_\beta \right \rangle =
\sum_\gamma
\left \langle \psi_\alpha \right . \left | \psi_\gamma \right \rangle
\left ( \phi_\alpha \right |
\hat{T}_B
\left | \phi_\beta \right )
=
\sum_\gamma
\left ( \phi_\alpha \right |
\hat{T}_B
\left | \phi_\gamma \right )
\left \langle \psi_\gamma \right . \left | \psi_\beta \right \rangle
\end{equation}
with $\hat{T}_B$ defined as above and
without any explicit reference to the norm operator, and similarly for a
two-body fermion operator $\hat{V}$.

Thus any boson representation of a Hamiltonian factorizes:
$\hat{\cal H}_B = \hat{\cal N}_B \hat{H}_B$.  Since the norm operator is a
function of the ${\rm SU}(2\Omega)$ Casimir operators it commutes with the
boson
images of fermion operators \cite{JG}, and one can simultaneously
diagonalize both $\hat{H}_B$ and $\hat{\cal N}_B$.  Then
Eqn.~(\ref{BosonEigen1}) becomes
\begin{equation}
 \hat{H}_B\left | \Phi_p \right ) =
\bar{E}_p \left | \Phi_p \right ).
\label{EigenBoson}
\end{equation}
where $\bar{E}_p = E_p$ for the physical states, but $\bar{E}_p$ for the
spurious states is no longer necessarily zero.

The boson Hamiltonian $\hat{H}_B$ is by construction Hermitian and, if one
starts with at most only two-body interactions between fermions, has at most
two-body
boson interactions.  All physical eigenstates of the original fermion
Hamiltonian will have counterparts
in (\ref{EigenBoson}).  It should  be clear that transition amplitudes
between physical eigenstates will be preserved. Spurious states will also
exist but, since the norm operator $\hat{\cal N}_B$ commutes with the boson
image
Hamiltonian $\hat{H}_B$, the physical eigenstates and the spurious states
will not admix. The spurious states can be identified because the norm
operator $\hat{\cal N}_B$ annihilates spurious states and  has
eigenvalue $(2N-1)!!$ on  physical states. To aid in the convergence to
the physical states in numerical computations we can put them low in the
spectrum by adding the
normal-ordered second order Casimir operator minus its eigenvalue for the
physical states,
\begin{equation}
\hat{M} = g \left[ \colon\hat{C}_2 \colon + 4N(N-1)\right]
\end{equation}
where $g$ is an arbitrary strength large enough to lower the physical states
below the spurious states.  Using the known eigenvalues of the Casimir
operators
for the irreducible representations of ${\rm SU}(2\Omega)$ \cite{Wybourne}, we
find that $\hat{M}$ has zero eigenvalue for
the physical states (those belonging to the totally antisymmetric irrep) and a
positive definite spectrum for the spurious
states.

We have mapped the fermion Hamiltonian into a finite,
Hermitian, boson Hamiltonian.  While diagonalizing this boson Hamiltonian
yields a spectrum that contains the eigenvalues of the original fermion
Hamiltonian, such a diagonalization would be numerically more taxing than the
original problem because of the additional spurious boson states.
Instead one wants to restrict the problem to a few boson degrees of freedom.
There are three steps in this process.

The first step is identifying the dominant degrees of freedom, either boson or
fermion-pair.   This may be done by assumption (prejudice), such as choosing
$J= 0,2$ fermion pairs or $s$- and $d$-bosons to match the IBM, or by some
variational method using  a  fermion-pair-condensate \cite{RoweVarPair} or
Hartree-Bose.

{}From a restricted set of fermion pairs, one can construct wavefunctions of
the form (2).  Diagonalizing the fermion Hamiltonian in this subspace yields
approximations to the true eigenstates; how good an approximation depends on
the first step.  The second step then is to find appropriate boson operators
for the  restricted space, that is operators that preserve the fermion matrix
elements and keep physical and spurious states from mixing, using only those
bosons corresponding to the allowed fermion pairs.
Construction of these operators cannot be carried out naively.
The boson representations of the
norm operator and of the Hamiltonian in this restricted space, which we
denote by $\left [ \hat{\cal N}_B \right ]_T$ and
$\left [ \hat{\cal H}_B \right ]_T$ respectively, are easily found:
one takes the boson representations in the full space (for the
norm operator  Eqn.~(\ref{NormCasimirRep}), (\ref{NormBosonRep}))
and keep only those terms consisting of allowed bosons.  The coefficients are
unchanged and matrix elements of states in the truncated fermion space are
still preserved.  Factorizing the Hamiltonian is not so simple, however, as
 $\left [ \hat{\cal H}_B \right ]_T
\neq   \left [ \hat{H}_B \right ]_T \left [ \hat{\cal N}_B \right ]_T$.
One {\it can}  make a factorization
\begin{equation}
\left [ \hat{\cal H}_B \right ]_T
=  \tilde{H}_T \left [ \hat{\cal N}_B \right ]_T;
\label{NewFactor}
\end{equation}
but $ \tilde{H}_T \neq \left [ \hat{H}_B \right ]_T$.
Instead one starts from a general ansatz which we simply sketch as
\begin{equation}
 \tilde{H}_T = \theta_1 b^\dagger b + \theta_2  b^\dagger b^\dagger b b
+ \theta_3  b^\dagger b^\dagger b^\dagger b b b + \ldots
\label{HTruncate}
\end{equation}
and find the coefficients $\theta_\ell$ (which depend on the truncation)
by normal-ordering the right-hand side of (\ref{NewFactor})
and equating term by term with the left-hand side.
The result is a boson image of the Hamiltonian whose spectrum contains the
eigenstates not of the full fermion space but of the restricted fermion space.

The final step, which is beyond the scope of this Letter to discuss in detail,
would be to renormalize the operators from the second step into effective
operators which account for the excluded degrees of freedom, producing
eigenvalues and transition amplitudes corresponding to the full fermion space.
If the assumptions of the first step are valid these corrections are small.

As an example we consider the pairing interaction.  For fermions with spin,
the dimension 2$\Omega$ is even and for, each fermion state $a^{\dagger}_i$,
there is a time-reversed state, $ K a^{\dagger}_i K^{\dagger} = \epsilon_i
a^{\dagger}_{\bar{\imath}}$ ($K$ is the time-reversal operator) where
$\epsilon_i = 1$ or $-1$ and $\epsilon_i =  -
\epsilon_{\bar{\imath}}$.  The paired two-particle state is then $\hat{A}_0
^{\dagger} = (4\Omega)^{-1/2} \sum_{i} \epsilon_i a
^{\dagger}_i a^{\dagger}_{\bar{i}}$ and hence $({\bf A}^{\dagger}_0)_{ij} =
(2\Omega)^{-1/2} \epsilon_i \delta_{j\bar{\imath}}.$  The pairing
interaction $\hat{P} = \hat{A}^\dagger_0 \hat{A}_0$  maps to the boson image
\begin{eqnarray}
  \hat{P}_B  =
{b}^\dagger_0 {b}_0
+ 2 \sum_{\sigma \sigma^\prime} \sum_{\tau \tau^\prime}
{\rm tr \,}
\left (
{\bf A}_\sigma {\bf A}^\dagger_0
{\bf A}_{\sigma^\prime} {\bf A}^\dagger_\tau
{\bf A}_0 {\bf A}^\dagger_{\tau^\prime}
\right )
{b}^\dagger_\sigma {b}^\dagger_{\sigma^\prime}
{b}_\tau {b}_{\tau^\prime}.
\end{eqnarray}
The boson image $\hat{P}_B$ can be solved completely and easily because,
like the original fermion Hamiltonian $\hat{P}$ \cite{TalmiText}, it
is a linear combination of Casimir operators,
\begin{equation}
\hat{P}_B = { 1 \over 4 \Omega} \left [
2 \hat{N} + \hat{C}_2 -2\hat{G}_2 \right ]
\end{equation}
where $\hat{G}_2$ is the Casimir operator of the symplectic subgroup
${\rm Sp}(2\Omega)$; it leaves the paired boson invariant,
$\left [ \hat{G}_2 , b^\dagger_0 \right ] = 0$.
Because the eigenstates of $\hat{P}_B$ must be simultaneously eigenstates of
$\hat{ \cal N}_B$, the naive guess
\begin{equation}
({b}^\dagger_0)^N  \left | 0 \right )
\label{NaiveGS}
\end{equation}
does not yield the ground state.
 This is not surprising since simple
correspondence, whereby the ground state of the pairing Hamiltonian maps into
$({b}^{\dagger}_0)^N  \left | 0 \right )$, leads to non-Hermitian Hamiltonians
\cite{Talmi}.
The norm operator $\hat{\cal N}_B$ acting on the ansatz (\ref{NaiveGS})
projects out the exact ground state; that projected wavefunction however
contains other bosons.

If our truncated space consists
of only fermion wavefunctions constructed from $\hat{A}^\dagger_0$ pairs,
and so in the boson space we have only $b_0^\dagger$-bosons, then
\begin{equation}
\left [ \hat{P}_B  \right ]_T= b^\dagger_0 b_0 + {1 \over 2 \Omega^2}
b^\dagger_0 b^\dagger_0 b_0 b_0.
\end{equation}
This has (\ref{NaiveGS}) as its only eigenstate but yields the wrong ground
state energy,
and one finds explicitly that
\begin{equation}
\left [ \hat{\cal P}_B \right ]_T \neq
\left [ \hat{P}_B \right ]_T  \left [  \hat{\cal N}_B \right ]_T .
\end{equation}
Instead,  the boson representation $ \left [\hat{\cal P}_B \right ]_T $
correctly factorizes to $ \tilde{P}_T  \left [  \hat{\cal N}_B \right ]_T $
with
\begin{equation}
\tilde{P}_T = b^\dagger_0 b_0 - {1 \over  \Omega}
b^\dagger_0 b^\dagger_0 b_0 b_0.
\end{equation}
This boson image, which is found starting from the ansatz (\ref{HTruncate}),
does indeed yield the correct ground state energy with (\ref{NaiveGS}) as its
eigenstate.

The factorization of a boson representation of a Hamiltonian
$\hat{\cal H}_B$ is guaranteed to yield a finite boson image only
in the full space; an arbitrary truncation with a general Hamiltonian
can induce higher-body terms, as allowed for in the
ansatz (\ref{HTruncate}).      The important question of the convergence
of the many-body terms induced by the truncation will be explored in the near
future.

In summary, the mapping of a fermion Hamiltonian into a boson Hamiltonian ,
where the bosons represent correlated fermion pairs, factorizes into a
finite boson Hamiltonian times a boson norm operator with an infinite number
of boson terms.  Because this finite boson Hamiltonian commutes with the
boson norm operator, the boson Hamiltonian does not mix physical and
spurious states.

This work was supported by the U.~S.~Department of Energy.

\pagebreak

\end{document}